\documentclass[aps,preprint,showpacs,prb]{revtex4}
\usepackage{times,xspace}
\usepackage{amsbsy,amssymb,amsmath,bm}
\usepackage{graphicx,color,epsfig,rotate}
\usepackage{fancyheadings}

\begin{document}
\title{LANDAU THEORY FOR SHAPE MEMORY POLYCRYSTALS}
\author{R. Ahluwalia}
\author{T. Lookman}
\author{A. Saxena}
\author{R.C. Albers}
\affiliation{Theoretical Division, Los
Alamos
National Laboratory,
Los Alamos, New Mexico, 87545 USA}
\date{\today}

\begin{abstract}
We propose a Ginzburg-Landau theory for the elastic properties of
shape memory polycrystals. A single crystal elastic free energy for a
system that undergoes a
square-to-rectangle transformation 
is generalized to a polycrystal by introducing
a crystal orientational field that is determined from a continuum phase 
field model.
The coupled system is used to study  domain morphology and 
mechanical properties of
shape memory alloys in different temperature regimes.
\\

\end{abstract}
\maketitle

\section{INTRODUCTION}
Materials known as martensites undergo first-order, diffusionless 
structural transformations from one crystal phase (austenite) to 
another, usually a twinned, phase (martensite) due to shear strains.  
A subclass of these materials with displacive transformations exhibits 
the shape memory effect \cite{otsu}. This effect refers to the existence 
of a residual strain upon unloading that can be recovered on heating to 
the high temperature austenite phase. In the austenite phase, these 
materials exhibit pseudoelastic behavior where a plateau in the stress-strain 
curves is observed; however, there is no residual strain as the macroscopic 
deformation is completely recovered when the load is removed.  Interestingly, 
these unusual mechanical properties of shape memory alloys do not involve 
any plastic effects such as those caused by dislocation motion but are entirely due 
to the intrinsic elastic nonlinearities. As a consequence of these unique 
properties, shape memory materials have many technological applications 
\cite{waym}.  Some examples include NiTi, FePd, AuCd and copper-based 
ternary alloys, e.g. CuAuZn. 

Most commercial applications of shape memory alloys make use of
polycrystalline specimens and therefore it is important 
to compare the mechanical response of polycrystals to that of single crystals.
The problem of finding the effective properties of
martensitic polycrystals has been studied by analytical methods
\cite{bhatta,brin,falk1} and finite element simulations \cite{Gall}. 
However, these methods do not account
for the complex polycrystal geometry and also do not
incorporate the long-range elastic interactions between the
grains.  Continuum simulations that span a range of length scales are good
candidates to describe these issues. Recently,
phase-field micro-elasticity models that employ static grains created by
the Voronoi construction have been studied 
\cite{khacha}.  
However, it is important to regard the grain orientation as a
thermodynamic  variable since a polycrystal, in reality, is a metastable 
state that is formed by a grain growth process. In the present work, we attempt to account for this metastability of a polycrystal.

The evolution of grains during grain
growth has been studied using the phase-field approach
\cite{elder1,chen,warren}. Although these models
correctly describe the grain morphologies and  domain  coarsening,
they do not address the issues of elasticity and
crystal symmetry. A coupling of these models with continuum elasticity 
models of martensitic transformations provides a framework to model 
mechanical properties of shape memory polycrystals.  Here we propose 
a model in which elastic strains are coupled to a phase-field model
through an orientation field that is determined from a
multi-component order parameter describing the crystal
orientations. Due to this coupling, the strains as well as
the grain orientations can change under an external load. In Section 2 
we describe our Ginzburg-Landau model and present two-dimensional 
($2D$) simulations of the loading-unloading cycle for martensitic 
polycrystals in different temperature regimes in Section 3. Our main findings 
are summarized in section 4.

\section{GINZBURG-LANDAU MODEL}
A Landau theory for shape memory materials was first proposed by Falk
\cite{falk2}. This one dimensional model captures the salient
physics of a number of experimentally observed features of
martensites. However, the model does not incorporate the elastic long-range
interactions that are crucial to describe the microstructure of martensites.
Barsch and Krumhansl introduced
a Ginzburg-Landau model to describe the inhomogeneous microstructure 
of these materials \cite{krum} and the  model has been extended to simulate
martensitic domain structures in two and three dimensions
\cite{shenoy,threed,lookman,jacobs,kartha}. Here
we generalize this theory to
describe a $2D$ square to rectangle martensitic transformation in a
polycrystal with different crystallographic orientations.
The theory is formulated in terms of a free-energy functional  
\begin{eqnarray}
F=\int d\vec{r}  \big[f_{grain} + f_{elastic} +f_{load}\big],
\end{eqnarray}
where $f_{grain}$ is the  free energy density due to the
orientational degrees of freedom of the polycrystal, $f_{elastic}$
represents the elastic free energy and $f_{load}$ is the free
energy contribution due to an external applied load. The
polycrystalline system is described by a set of $Q$ non-conserved
order parameters \cite{chen}$(\eta_{1},\eta_{2},...,\eta_{Q})$. 
In
terms of these order parameters the free energy $f_{grain}$ is
given by 
\begin{equation}
f_{grain}=\sum^{Q}_{i=1}\bigg[{{a_1}\over{2}}{\eta_i}^2
+{{a_2}\over{3}}{\eta_i}^3 +{{a_3}\over{4}}{\eta_i}^4\bigg]+
{{a_4}\over{2}} \sum^{Q}_{i=1} \sum_{j>i}{\eta_i}^2 {\eta_j}^2
+\sum^{Q}_{i=1} {{K}\over{2}}(\nabla \eta_i)^{2}.
\end{equation}
For $a_1,a_2 < 0$ and $a_3,a_4 > 0$, the first two terms in
equation (2) describe a potential with $Q$ degenerate minima 
$(\eta_0,0,...,0)$, $(0,\eta_0, ....,0)$ up to $(0,0,....,\eta_0)$ 
$(\eta_0 > 0)$,
corresponding to $Q$ grain orientations.  This form differs from that 
used by Chen and Yang \cite{chen} as there is an additional cubic term 
in the free energy. The cubic term ensures that ${\eta_0}\ge 0$ and 
allows us to uniquely associate an orientation with each minimum,
given by an angle 
\begin{equation}
\theta(\vec{\eta},\vec{r})={{\theta_m}\over{Q-1}}
\bigg[{{\sum^{Q}_{i=1} i{\eta_i}(\vec{r})}\over {\sum^{Q}_{i=1}
{\eta_i}(\vec{r})}}-1\bigg].
\end{equation}
Thus, there are $Q$ orientations between $0$ and a maximum angle 
$\theta_{m}$ (as an example, the $Q=3$ case has minima at 
$(\eta_0,0,0)$ and  
$(0,\eta_0,0)$ and  
$(0,0,\eta_0)$ corresponding to 
$\theta=0^{o}$, 
$\theta={\theta_m}/2$ and  
$\theta={\theta_m}$). 

The gradient energy $(K >
0)$ represents the energy cost of creating a grain boundary.
To describe elastic effects, we consider the linearized strain tensor 
in a global reference frame defined as $\epsilon_{ij}=(u_{ij}+u_{ji})/2$ 
$(i=1,2; j=1,2)$, where $u_i$ represents a component of the
displacement vector and $u_{ij}$ is a displacement gradient. We use 
the symmetry-adapted linear combinations of the strain tensor \cite{krum}
defined by $\epsilon_1=(\epsilon_{xx}+\epsilon_{yy})/\sqrt2$, $\epsilon_2 
=(\epsilon_{xx}-\epsilon_{yy})/\sqrt2$ and $\epsilon_3=\epsilon_{xy}$. 
Here $\epsilon_1$ represents the bulk (dilatation) strain, $\epsilon_2$ 
the deviatoric (rectangular) strain and $\epsilon_3$ the shear strain. To 
generalize these definitions to the case of a polycrystal described
by an orientational field $\theta(\vec{\eta})$, the strain tensor 
in a rotated frame is calculated using $R(\theta(\vec{\eta}))\stackrel 
{\leftrightarrow}{\epsilon}R^{T}(\theta(\vec{\eta}))$, where 
$R(\theta(\vec{\eta}))$ is a rotation matrix and $\theta(\vec{\eta})$ is 
determined from the minima of $F_{grain}$ using equation (3). Under this 
rotation, the elastic free energy describing a square to rectangle 
transition is given by 
\begin{equation}
f_{elastic}={{A_1}\over{2}}{e_1}^2 +{{A_3}\over{2}}{e_3}^2
+f_{local}(e_2)+
{{K_2}\over{2}}{(\nabla e_2)}^2,
\end{equation}
 where $e_1=\epsilon_1$,
$e_2={\epsilon_2}\cos[2\theta(\vec{\eta})]+\sqrt{2}{\epsilon_3}\sin[2\theta(\vec{\eta})]$
and
$e_3=-(1/\sqrt{2}){\epsilon_2}\sin[2\theta(\vec{\eta})]+
{\epsilon_3}\cos[2\theta(\vec{\eta})]$. The nonlinear part of the 
elastic free energy is 
\begin{equation}
f_{local}(e_2)={{A_2}\over{2}}{e_2}^2+
{{\alpha}\over{4}}{e_2}^4+
{{\beta}\over{6}}{e_2}^6.
\end{equation}
At the level of the unit cell, the transformation free energy $f_{local}$ 
describes a square to rectangle
transition where the austenite phase (square) has ${e_2}=0$ and 
the martensite
phase is described by ${e_2}=\pm e_0$ corresponding to the two 
rectangular variants.
Here $A_1=C_{11}+C_{12}$, $A_2=C_{11}-C_{12}$ and $A_3=4C_{44}$,
where $C_{11}$, $C_{12}$  and $C_{44}$ are the elastic constants
for a crystal with square symmetry. The quantities $\alpha$ and $\beta$ are the higher order nonlinear elastic 
constants and $K_2$ is the
appropriate deviatoric strain gradient coefficient that 
determines the energy cost of creating a domain wall (twin boundary) between the two rectangular variants. In principle, $K_2$ can be
determined experimentally from phonon dispersion curves \cite{kartha}.
 In this work, we are interested in simulating a
uniaxial loading experiment. If we choose the $x$ axis to be the
loading axis, then the free energy contribution due to the
external load $\sigma$ is given by 
\begin{eqnarray}
f_{load}&=&-\sigma \epsilon_{xx} 
=-{{\sigma}\over{\sqrt{2}}} (\epsilon_{1}+\epsilon_{2})
\nonumber \\ &=&-{{\sigma}\over{\sqrt{2}}}
\bigg[e_{1}+{e_2}\cos[2\theta(\vec{\eta})]
-\sqrt{2}{e_{3}}\sin[2\theta(\vec{\eta})]\bigg].
\end{eqnarray}
The strains $\epsilon_{1}$
, $\epsilon_{2}$
and $\epsilon_{3}$ are not independent but are related by the 
elastic compatibility
relation \cite{love} 
\begin{eqnarray}
{{\nabla}^2}{\epsilon_1}-({{\partial^2}\over{\partial x^2}}
-{{\partial^2}\over{\partial y^2}}){\epsilon_2}-\sqrt{8}
{{\partial^2}\over{ {\partial x}{\partial y} }}{\epsilon_3}=0.
\end{eqnarray} 
In terms of the strain variables $e_1$, $e_2$ and $e_3$, this relation becomes
\begin{eqnarray}
{{\nabla}^2}{e_1}&-&({{\partial^2}\over{\partial x^2}}
-{{\partial^2}\over{\partial
y^2}})\bigg[{e_2}\cos[2\theta(\vec{\eta})]-\sqrt{2}{e_3}\sin[2\theta(\vec{\eta})]\bigg]\nonumber\\
&-&\sqrt{8} {{\partial^2}\over{ {\partial x}{\partial y}
}}\bigg[{{e_2}\over{\sqrt{2}}}\sin[2\theta(\vec{\eta})]+
{e_3}\cos[2\theta(\vec{\eta})]\bigg]=0.
\end{eqnarray}
We eliminate the strain $e_1$ from $f_{elastic}$ and $f_{load}$ using 
equation (8)
so that the total elastic energy is in terms of $e_2$ and $e_3$ only.
This method has  been used previously in the context of martensitic
transformations in single crystals\cite{lookman}. We first 
introduce ${e'_1}=e_1 -(\sigma/A_{1}\sqrt{2})$ so that
$f_{eff}=f_{elastic}+f_{load}$, where
\begin{eqnarray}
f_{eff}&=&{{A_1}\over{2}}{{e_1}^{'}}^2 +{{A_3}\over{2}}{e_3}^2
+f_{local}(e_2)
-{{\sigma}\over{\sqrt{2}}}
\bigg[{e_2}\cos[2\theta(\vec{\eta})]
-\sqrt{2}{e_3}\sin[2\theta(\vec{\eta})]\bigg].
\end{eqnarray}
To eliminate ${e'_{1}}$ from equation (9), we use the Fourier representation
of equation (8) to obtain 
\begin{equation}
{e'_1}(\vec{k})={ {{k_x}^2-{k_y}^2}\over{{k_x}^2+{k_y}^2}}\Gamma_2(\vec{k})
+{ {\sqrt{8}{k_x}{k_y}}\over{{k_x}^2+{k_y}^2}}\Gamma_3(\vec{k}),
\end{equation}
where $\Gamma_2(\vec{k})$ represents the Fourier transform of
${e_2}\cos[2\theta(\vec{\eta})]-\sqrt{2}{e_3}\sin[2\theta(\vec{\eta})]$
and $\Gamma_3(\vec{k})$ is the Fourier transform of
${e_2}(\sin[2\theta(\vec{\eta})]/\sqrt{2})+{e_3}\cos[2\theta(\vec{\eta})]$.
Thus, the effective free energy can be written as
\begin{eqnarray}
F_{eff}&=&{{A_1}\over{2}}\int d\vec{k}\bigg[ \bigg({
{{k_x}^2-{k_y}^2 }\over{ {k_x}^2+{k_y}^{2} }
}\bigg)^{2}|\Gamma_2(\vec{k})|^{2} +\bigg({ {{\sqrt{8}{k_x}{k_y} }
\over{ {k_x}^2+{k_y}^{2} }
}\bigg)^{2}}|\Gamma_3(\vec{k})|^{2}\nonumber\\ &+&{
{{\sqrt{8}{k_x}{k_y}({{k_x}^2-{k_y}^2)} } \over{
({k_x}^2+{k_y}^{2})^{2} } }} \bigg(\Gamma_3(\vec{k})
\Gamma_2(-\vec{k})+ \Gamma_3(-\vec{k})
\Gamma_2(\vec{k})\bigg)\bigg] \nonumber \\ &+&\int
d\vec{r}\bigg[{{A_3}\over{2}}{e_3}^2
+f_{local}(e_2)
-{ {\sigma}\over{\sqrt{2}} }
\bigg({e_2}\cos[2\theta(\vec{\eta})]
-\sqrt{2}{e_3}\sin[2\theta(\vec{\eta})]\bigg)\bigg]. 
\end{eqnarray}
The long-range part of the free energy 
is always orientation dependent due to the elastic compatibility induced
anisotropy in the kernels. This  term ensures that elastic
compatibility will be satisfied within the grains as well as at the
grain boundaries. We emphasize that the long-range interaction excludes the $|\vec{k}|=0$ mode 
since compatibility is trivially satisfied for this case. The $|\vec{k}|=0$ 
mode, which refers to   
the homogeneous 
state, is accounted for by the local terms in $F_{eff}$. 
The total free energy of the system is  
$F=F_{grain}+F_{eff}$, and we assume relaxational dynamics for 
$e_2$ and $e_3$, that is,
\begin{eqnarray}
{{ \partial e_2} \over {\partial t} }
&=&-{\gamma_2}\left[{{\delta F}\over{\delta e_2} }
\right], \nonumber\\
{{ \partial e_3} \over {\partial t} }
&=&-{\gamma_3}\left[{{\delta F}\over{\delta e_3} }
\right],
\end{eqnarray}
where $\gamma_{2}$ and $\gamma_{3}$ are the appropriate kinetic
coefficients for the deviatoric and shear strains.
Similarly, the dynamics of the grains is defined by $Q$
 equations 
\begin{eqnarray}
{{\partial \eta_i}\over{\partial t}}=-
{\gamma_{\eta}}{{\delta F}\over{\delta \eta_i}},
\end{eqnarray}
where $\gamma_{\eta}$ is a kinetic coefficient and $i=1,...,Q$,
corresponding to $Q$ grain orientations.

\section{SIMULATIONS OF TEXTURE AND STRAIN EVOLUTION} 
We simulate the mechanical properties of shape memory
materials using the model described in Section 2. We choose 
FePd parameters \cite{kartha} 
for which $A_1=140$ GPa, $A_3=280$ GPa, $\alpha=-1.7\times{10^4}$ GPa 
and $\beta=3\times{10^7}$ GPa. The temperature dependent elastic
constant $A_2$ undergoes a softening and hence controls
the square to rectangle transformation.  We study four different cases
corresponding to $A_2=-2,1,2,3$ GPa. Muto et al. \cite{muto} have measured 
the elastic constants of FePd as a function of temperature. The lowest 
temperature measurement they reported was at $290$ K corresponding to $A_2 
\sim 10$ GPa. Thus the values of $ A_2$ we have chosen correspond to 
temperatures lower than $290$ K and are in the vicinity of the transition 
temperature of $265$ K. Figure 1 shows the profiles for the free energy 
$f_{local}$ for these values. For the parameters in $f_{grain}$ we choose 
(for illustrative purposes) $a_1=-10$ GPa, $a_2=-10$ GPa,
$a_3=10$ GPa, $a_4=20$ GPa, $Q=5$ and ${\theta_m}=30^o$. Here, we also
need to specify the grain boundary energy coefficient $K$ and
the strain gradient coefficient $K_2$. For FePd, the strain
gradient coefficient \cite{kartha} has been measured to be ${K_2}/{a_0}^2=25$ 
GPa,
where $a_0$ is the lattice spacing of the crystal. The grain
boundary energy coefficient is chosen as $K/{a_0}^2={10^{5}}$
GPa. The lengths  are scaled by $\vec{r}=(100{a_0})\vec{\zeta}$.
For a homogeneous single crystal, using
these parameter values, the
free energy in equation (2) has $5$ degenerate minima defined by
$\theta_0(\vec{\eta})={0^o}, {7.5^o},{15^o},{22.5^o},30^o$. We should point 
out that we  
can consider a more continuous orientation distribution of the polycrystal with a large number of states (large value of $Q$) 
in the full range ($0^{o}-45^{o}$). However,  for the sake of clarity of the domain patterns, we restrict ourselves to the range described above.
Equations (12) and (13) are solved numerically to simulate the
domain structures and mechanical properties in different regimes.
For simplicity, we assume
$\gamma_{\eta}=\gamma_{2}=\gamma_{3}=\gamma$ and use rescaled time
defined by $t^*=t({10^{10}}{\gamma})$.

An initial
polycrystalline configuration is first generated by solving equations (12) and
(13) for a $12800{a_0}\times 12800{a_0}$ system with periodic boundary conditions, starting from random initial conditions. A grain growth process is simulated
with $\sigma=0$ in the austenite phase so that all components of the strain 
tensor vanish.  Grains
with orientations $\theta_0(\vec{\eta})={0^o},
{7.5^o},{15^o},{22.5^o},30^o$ form and start coarsening. We arrest
the system in a given polycrystalline
configuration by abruptly changing the value of the parameter
$a_1$ from $-10$ GPa to $-160$ GPa (the parameter $A_2$ is also
changed  so that the system is in the desired martensitic phase). This
sudden decrease in $a_1$ increases the free energy barriers between
the crystalline states and the growth stops.
We first consider the case $A_2=-2$ GPa. Figure 1 shows that for a 
homogeneous system, the 
transformation free energy $f_{local}({e_2})$ for $A_2=-2$ GPa has a local maximum 
at $e_2=0$ and two degenerate global minima. In the absence of
applied stress ($\sigma=0$), the arrested polycrystal evolves into a domain
pattern of the variants of the martensitic phase (there is no austenite present 
since $e_2=0$ is unstable). The strains in each grain as well as the
orientation of the martensitic domain walls (i.e. twin boundaries) are determined by the orientation of the
grain. This behavior is clear from Figure 2(a) that shows the distribution of
the strain ${\epsilon_2}(\vec{r})$ (deviatoric strain relative to the global
frame of reference) and Figure 2(b) that shows the local orientation
$\theta(\vec{\eta}(\vec{r}))$. The domain walls are oriented at angles
$\theta(\vec{\eta})+\pi/4$ or $\theta(\vec{\eta})-\pi/4$. We point out
that the average strains for this configuration are
very small and correspond to a system with no macroscopic deformation.

To simulate mechanical loading, an external
 tensile stress $\sigma$ is applied quasi-statically, i.e., starting from the
unstressed
configuration of Figures 2(a) and 2(b), the applied stress $\sigma$ is
increased in steps of $ 5.13$ MPa, after allowing the configurations to relax
for $t^{*}=25$ time steps after each increment. The loading is continued till a
maximum stress of $\sigma=200$ MPa is reached.
Thereafter, the system is unloaded by decreasing $\sigma$ to zero at the
same rate at which it was loaded. Figures 2(c) and 2(d) relate to  
a stress level of $\sigma=46.15$ MPa during the loading process. 
The
favored variants (red domains in the left panel) have started to grow at the expense of the
unfavored variants (blue domains in the left panel). The orientation distribution
$\theta(\vec{\eta}(\vec{r}))$ in
Figure 2(d) has not changed much. As the stress level is increased further, the
favored variants  grow.  Even at the maximum stress of $200$ MPa,
some unfavored variants persist, as is clear from Figure 2(e) (in fact, further 
application of stress does not remove such structures as the unfavored 
variants are in a ``locked'' state due to intergranular constraints). 
We note that the grains with large misorientation with the loading direction 
rotate, as is clear by comparing Figure 2(f) with Figure 2(b)
and Figure 2(d), where we can observe that dark red colored grains have turned
orange, indicating rotation of those grains. Grains with lower
misorientation do not undergo significant rotation. This
rotation is due to the tendency of the system to maximize the transformation strain
in the direction of loading so that the total free energy is minimized.
Within the grains that rotate, sub-grain bands with slightly higher values
of the orientation $\theta(\vec{\eta}(\vec{r}))$ are present. These bands
correspond to the unfavored strain variants that still survive. Figure 2(g)
 and 2(h) show the situation after unloading to $\sigma=0$.
Upon removing the load, a domain structure is nucleated again due to the local
strain  gradients at the grain boundaries and the surviving unfavored variants
in the
loaded polycrystal configuration in Figure 2(e). This domain structure is
not the same as that
prior to loading (Figure 2(a)) and thus there is an underlying hysteresis. The
unloaded configuration has non-zero average strain. This average strain
is recovered by heating to the austenite
phase, as per the shape memory effect.
Figure 2(h) shows that the
orientation distribution reverts to its preloading state as the grains rotate
back when the load is removed. To show the rotation of grains more clearly, we 
plot in Figure 3 the change in orientation $\Delta \theta=\theta(\sigma)-\theta(\sigma=0)$
for $\sigma=200$ MPa. It is clear by comparing this figure with Figure 2(b) that 
only grains 
with large misorientations with the loading axis ($\theta=0^{o}$)  
rotate in order to decrease the misorientation. In some of these rotating grains, we observe bands where the misorientation has increased. As discussed earlier, 
these bands correspond to regions where the unfavored variants do not
disappear even at high stresses. This behavior is a consequence of the intergranular constraints.  

We compare the above  mechanical behavior of
the polycrystal to the corresponding single crystal. A single
crystal simulation is set up with exactly the same elastic free
energy parameters as the polycrystal case but the orientation is
fixed at $\theta=0^{o}$. Loading conditions are also identical to
the polycrystal case. Figure 4 shows the evolution of the
martensitic variants during the loading-unloading cycle for the
single crystal. Figure 4(a), shows the simulated
microstructure ($12800{a_0}\times 12800{a_0}$) prior to loading,
with  domain wall orientations at $\pi/4$
or $-\pi/4$ everywhere. In contrast, the domain wall orientations
change from grain to grain in the polycrystal case depicted in
Figure 2(a). Figure 4(b) shows the domain patterns for a stress
$\sigma=46.15$ MPa. We see that the favored variants (red
domains) grow at the expense of the unfavored
variants (blue domains). This growth continues, as shown in Figure 4(c) ($\sigma=56.41$ MPa)
and Figure 4(d) ($\sigma=71.79$ MPa). 
Finally,   in Figure 4(e)
($\sigma=200$ MPa) all unfavored variants disappear and we obtain a
single domain, in contrast to the
polycrystal case at the same stress level (Figure 2(e)) where
unfavored variants persist. Upon unloading, the single crystal
remains in a single domain state (Figure 4(f)). This is because 
there are no inhomogeneities or thermal noise in the simulations to cause 
renucleation of the domain structure and so the system remains in the
positive strain minimum homogeneously. The unloaded
polycrystal, however, reverts to the domain pattern
(Figure 2(g)). This also shows that, unlike the single crystal case, the 
mechanical behavior of the polycrystal is not necessarily 
governed by  $f_{local}(e_2)$ but by 
a more complex inhomogeneous free energy landscape.
A consequence of this is that the residual strain for a single crystal
\cite{nial} will be larger than that for a polycrystal \cite{nial1}.
The stress-strain curves corresponding to Figure 2 and Figure 4
are shown in Figure 5. The residual strain for the polycrystal ($\sim0.7\%$) 
is smaller than that for the single crystal ($\sim1.8\%$) due to an 
effective averaging over different orientations and 
nucleation of 
domains at grain boundaries upon unloading. Also, the change in the 
stress-strain curve for the polycrystal is not abrupt because the
response of the polycrystal is averaged over all grain orientations.

We now discuss cases where the material exists in the austenite phase prior 
to loading. As can be seen in Figure 1, $A_2=1,2$ and $3$ GPa 
correspond to this 
situation.
For $A_2=1$ GPa, the austenite $(e_2=0)$ is a metastable 
local minimum and there are two stable martensitic minima.
Figure 6(a) and 6(b) show the situation for $\sigma=0$. Since there are no 
nucleation mechanisms in the present simulations, the system remains in an 
austenite 
phase $(e_2=0)$ in all the grains. Loading is simulated in exactly
the same manner as for the $A_2=-2$ case. In Figure 6(c) and Figure 6(d) we  
observe that a stress induced transformation has heterogeneously 
occured in all the grains ($\sigma=30.76$ MPa) (although no local stresses 
have been introduced in the model, the heterogeneous nucleation occurs due to 
the strain gradients at the grain boundaries which act as embryos for the 
transformation). 
The 
transformed regions are represented by yellow/red shades whereas the 
untransformed 
austenite regions are represented by light green shades. Some grains also show 
the presence of the unfavored variant $(e_2 < 0)$ represented by dark green and
blue shades. Figure 6(d) shows that at this stress level, the 
texture is the same as for the unloaded configuration. At the 
maximum stress (Figure 6(e))
$\sigma=200$ MPa, most of the system has 
transformed although some austenite and the unfavored variant domains remain. 
As in the earlier 
case, grains with higher misorientation have also rotated (Figure 6(f)). 
Upon unloading, for this 
case
also some of the unfavored domains reappear, as can be seen in Figure 6(g). 
Figure 6(h) 
shows that the texture  
returns to its preloading state.

Next, we consider the case $A_2=2$ GPa for which $e_2=0$ is the global minimum and there are two 
metastable martensitic minima. For this case, the
``arrested" polycrystal exists in the austenite phase before loading.
This situation is depicted in Figure 7(a) where the strain
$\epsilon_{2}$ is close to zero everywhere. The corresponding orientation
distribution is shown in Figure 7(b). The system is loaded at an 
identical loading rate as in the earlier cases.
 Figure 7(c) shows the spatial distribution of
$\epsilon_{2}$ at a load of $\sigma=30.76$ MPa. A {\it stress induced}
martensitic transformation from the square phase to the
rectangular phase takes place heterogeneously, as can be seen from
the transformed regions (yellow/red shades) that are embedded in a
matrix of the untransformed austenite (green/blue shades). 
 Figure 7(d) shows that at this
stress level, the orientation distribution does not change
appreciably. Upon loading further to $\sigma=200$ MPa, the transformed
phase grows although the austenite phase is locally retained in
some regions (Figure 7(e)). Similar to the earlier cases,
transformation accommodating grain rotations are also observed for
this case, as seen in Figure 7(f). When the load is removed, the
domain structure shown in Figure 7(g) is observed. Here, most of
the system reverts to an austenite phase but some martensitic domains
 (as yellow streakings) remain giving rise to a very small 
residual strain. Figure 7(h)
shows the corresponding orientation distribution that returns to its
preloading state.

Finally, we discuss the case $A_2=3$ GPa which delineates a pseudoelastic behavior. The appropriate 
free energy shown in Figure 1
 has only one minimum, corresponding to $e_2=0$. For this case as well, the
preloading state is austenite and the strain $\epsilon_{2}$ is
very small everywhere, as seen in Figure 8(a). The corresponding
orientation distribution is shown in Figure 8(b). Nucleation of
martensitic variants in the austenite matrix can be observed in Figure
8(c), corresponding to $\sigma=30.76$. Figure 8(d) shows that the 
orientation distribution associated with this stress does not
change much. As the stress is further increased to $\sigma=200$ MPa, most
of the system is transformed to (stress induced) martensite, as shown in Figure
8(e). Rotations of grains with large misorientation with the
loading axis are observed for this case also (Figure 8(f)). In
contrast to the $A_2=-2$ GPa and $A_2=2$ GPa cases, this system
reverts to a homogeneous austenite phase upon
unloading (Figure 8(g)), i.e. there is no residual strain. Thus,
this case shows pseudoelasticity. The grain rotations are also
recovered upon unloading, as seen in Figure 8(h).

We have also simulated the single crystal loading-unloading cycle corresponding to 
Figures 6, 7 and 8. 
Unlike the $A_2=-2$ GPa case discussed in Figure 3, the system always remains in a single domain 
state during the loading and unloading.  This is because there are no nucleation mechanisms 
in the single crystal cases to create twin boundaries.
A stress induced martensitic transformation occurs homogeneously 
at a critical load, for all these cases. Upon unloading, there is no renucleation of the twinned state.

The stress-strain curves corresponding to Figures 6, 7 and 8 are depicted  
in Figure 9. The corresponding single 
crystal curves are also shown. 
It is clear that the residual strain decreases as the elastic constant $A_2$ (or the temperature) 
increases. It is also seen 
that the residual strains for the polycrystal simulations corresponding to $A_2=1$ GPa and $A_2=2$ GPa are 
smaller than their single crystal counterparts due to averaging over 
different orientations and
nucleation mechanisms that exist in the polycrystal. 
However, for the pseudoelastic case ($A_2=3$ GPa), the residual strains are zero for both the single- and 
polycrystal cases.
\section{Summary and Discussion}
We have proposed a framework to study the
mechanical properties of shape memory polycrystals. We have coupled 
the  elastic free energy for a square to rectangle transition 
to a phase-field model describing crystal 
orientations.
This approach can
be readily extended to any crystal symmetry and does not require
any a priori assumption of grain shapes or microstructure. The
microstructure is governed by the crystal symmetries encoded in
the appropriate elastic free energy functional. The long-range elastic
interaction between the grains is also incorporated. An important
feature of the present work is the coupling between the grain
orientation and elasticity so that
the metastability of the polycrystal may be accounted for  
within the same framework.

We studied 
mechanical properties of shape memory polycrystals and single crystals 
in different temperature regimes. There are significant differences between 
the mechanical 
response of single- and polycrystals. Since the mechanical properties of the
polycrystal
are an average of individual grains, the stress-strain curves are smoother 
compared to those of the  
single crystals. The inhomogeneities in the polycrystal ensure that
domain walls influence the mechanical behavior throughout the loading-unloading
process. In the temperature regimes with nonzero residual strain, the unloaded
polycrystals not only have reduced strain but also show domain microstructure.  
In contrast, the simulated defect free
single crystals 
exhibit no such patterns after unloading and have much
higher residual  
strains. Our findings are consistent with the fact that in general, polycrystals have poor 
shape memory properties in comparison to single 
crystals \cite{bhatta,nial,nial1}. We emphasize that our findings apply to 
materials where the symmetry change due to the transformation is not 
particularly large and the number of low symmetry variants is relatively
small (for example materials that show a cubic to tetragonal transformation). 
Materials such as NiTi, which undergo a cubic to monoclinic transformation
 have a 
relatively larger symmetry change and larger number of low symmetry variants. 
For NiTi, the differences between shape memory behavior of polycrystals and 
single crystals is not significant. Thus, the issues of how well strain is 
accommodated and how favored are the variants can play a role in determining the shape memory properties.
 Clearly, further investigation of this aspect is required.

The simulations also predict that grains with a large misorientation
with the loading direction 
can undergo reversible rotations, even in the absence of plastic flow. The magnitude of the rotation is  very small  
in the linear elastic regime and significant rotation occurs only after the transformation 
 begins.
It is well 
known that grains can rotate in the plastic
regime to accommodate crystallographic slip \cite{poulsen}. Although there are
no plastic effects within the present model, there are elastic nonlinearities that
couple to the orientation degrees of freedom. Measurements of texture
evolution upon loading in 
NiTi shape memory alloys have recently been reported \cite{vaidya}. However,
from these experiments, it is not possible to clarify if the rotational 
mechanism we observed exists.  
Our 
simulations
suggest that for grains with a high misorientation with the loading axis , in principle, rotation as well as detwinning can occur simultaneously for
shape memory polycrystals.  There appears to be an underlying tendency in the
polycrystal to decrease the effect of misorientations. Based on experiments and finite element simulations, a similar tendency where favorably oriented grains induce the transformation in unfavorably oriented grains, thereby effectively 
reducing the effect 
of grain misorientations, has been discussed by Gall et al \cite{Gall}. Since 
the grain orientations are not fixed in our simulations, this 
reduction in the effect of misorientations can be achieved by
evolution of the transformation strains as well as the grain orientations.
Further experiments  are needed to clarify
 to what extent such a mechanism indeed exists in real shape memory 
polycrystals.
The magnitude of this grain rotation depends on the parameters of
the polycrystal phase field model, or on the energy barrier between the grains. 
These parameters have been chosen 
arbitrarily for illustrative purposes in the present simulations.    
An estimation of these parameters from experiment or
atomistic simulations will be useful for a better comparison with experiments.  
\section*{ACKNOWLEDGEMENT}
We are grateful to F. L. Addessio and T. O. Williams for fruitful discussions. 
This work was supported by the U.S. Department of Energy.

\newpage
\begin{figure}[h]
\epsfysize=7cm
\epsffile{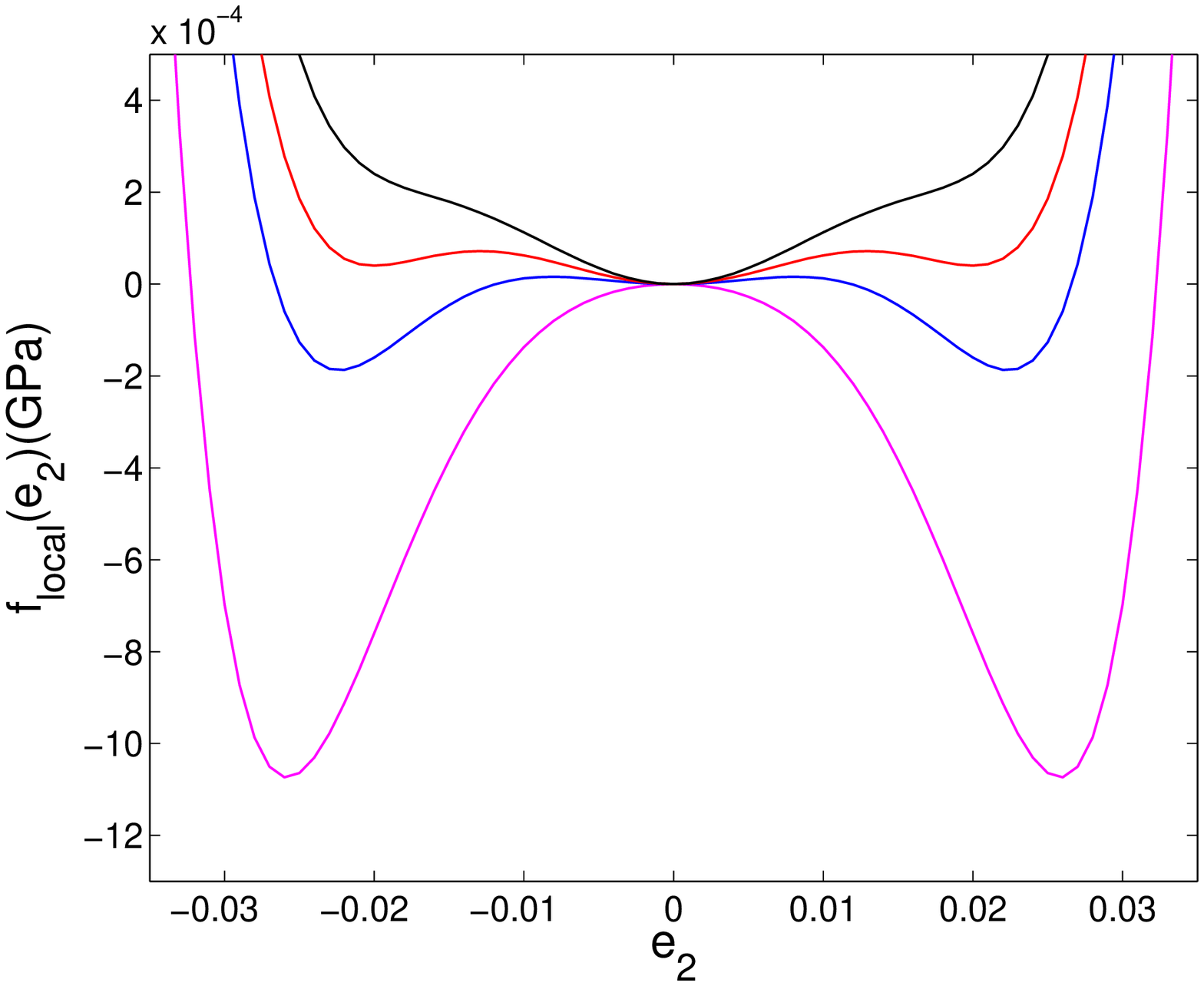}
\caption{
Local free energy density vs. $e_2$ as a function of temperature. The relatively high temperature 
(black curve) has  $A_2=3$ GPa, and successive curves with decreasing temperatures are for $A_2=2,1$ 
and $-2$ GPa, respectively.}
\label{fig1}
\end{figure}
\newpage
\begin{figure}[h]
\epsfysize=21cm
\epsffile{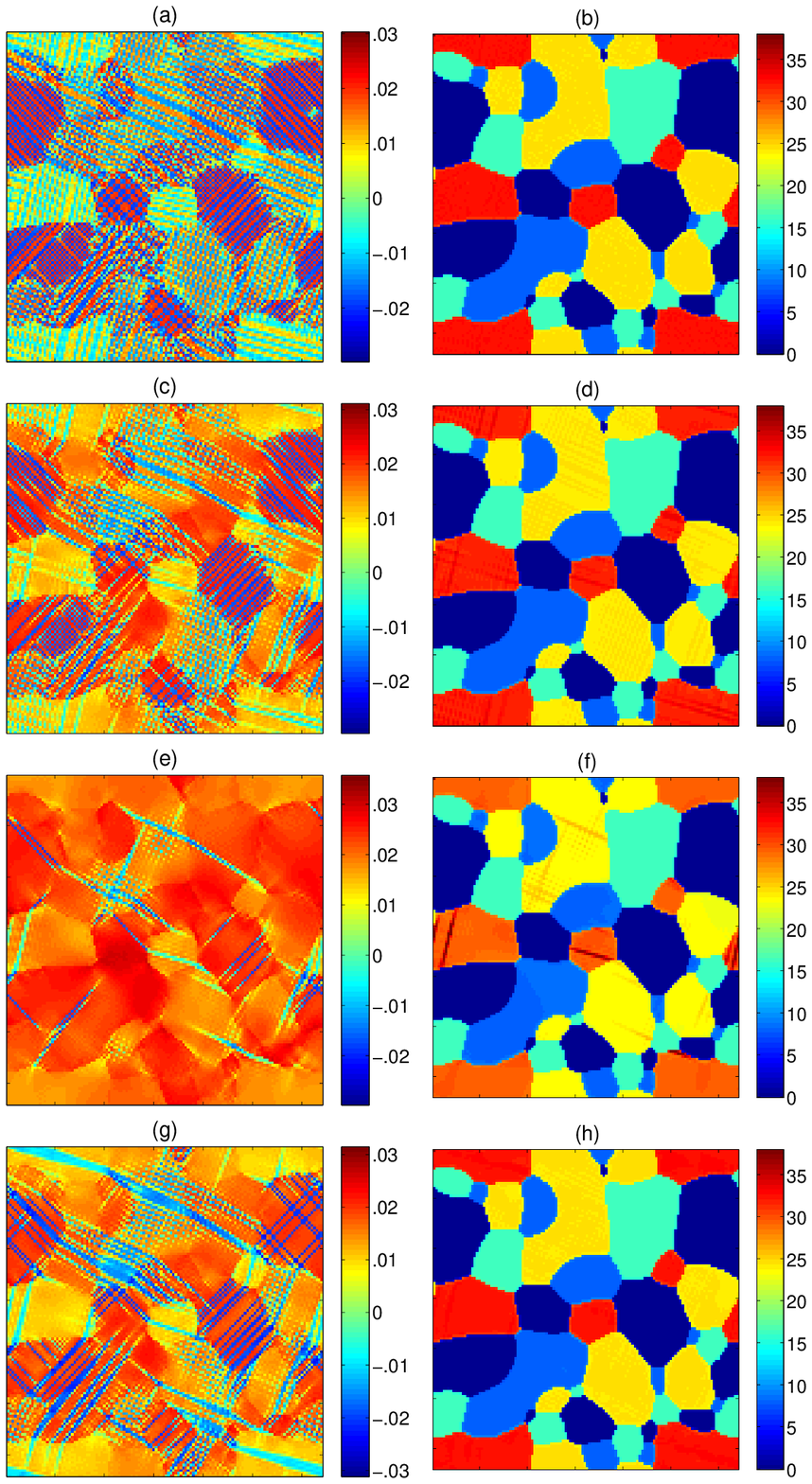}
\caption{
Spatial distribution of $\epsilon_2(\vec{r})$ 
(deviatoric strain in a global frame) for a polycrystal with $A_2=-2$ GPa
(snapshots (a),(c),(e) and (g)) and $\theta(\vec{r})$ in degrees
(snapshots (b),(d),(f) and (h)). The corresponding stress levels are
$\sigma=0$ (before loading, (a) and (b)),
$\sigma=46.15$ MPa ((c) and (d)),
$\sigma=200$ MPa ((e) and (f)) and
$\sigma=0$ (after unloading, (g) and (h)). 
}
\label{fig2}
\end{figure}
\newpage
\begin{figure}[h]
\epsfysize=3.5cm
\epsffile{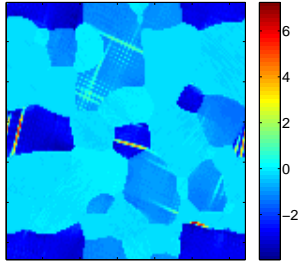}
\caption{
Spatial distribution of $\Delta \theta=\theta(\sigma)
-\theta(\sigma=0)$ in degrees for $\sigma=200$ MPa (corresponding to Figure 2(f)).} 
\label{fig1}
\end{figure}
\newpage
\begin{figure}[h]
\epsfysize=21cm
\epsffile{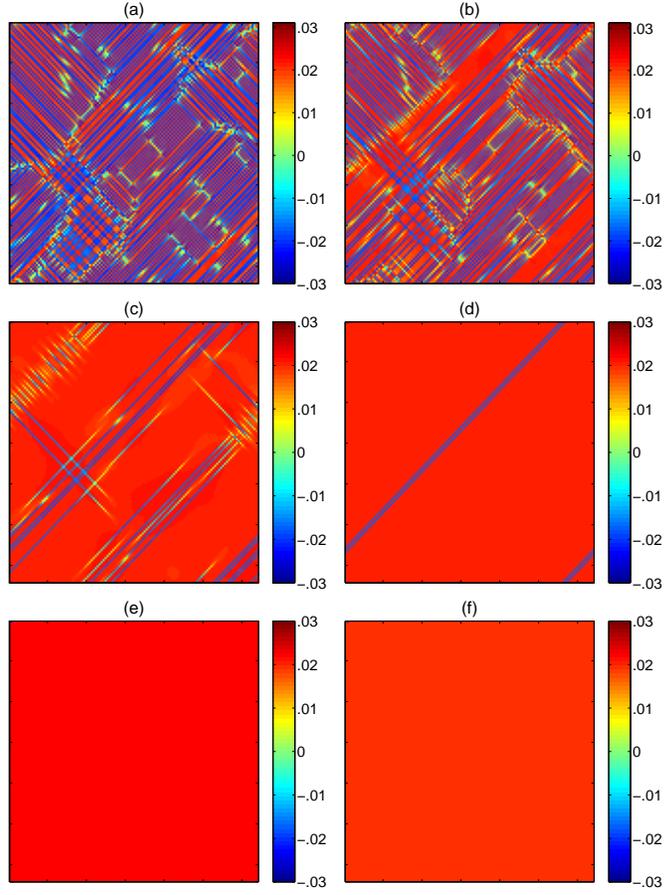}
\caption{
Spatial distribution of $\epsilon_2(\vec{r})$ 
(deviatoric strain in a global frame) for a single crystal 
$(\theta(\vec{\eta})=0
^{o}$) 
with $A_2=-2$ GPa.
 The corresponding stress levels are
$\sigma=0$ (before loading) (a),
$\sigma=46.15$ MPa (b),
$\sigma=56.41$ MPa (c),
$\sigma=71.79$ MPa (d),
$\sigma=200$ MPa (e)
 and $\sigma=0$ (after unloading) (f).
}
\label{fig1}
\end{figure}
\newpage
\begin{figure}[h]
\epsfysize=7cm
\epsffile{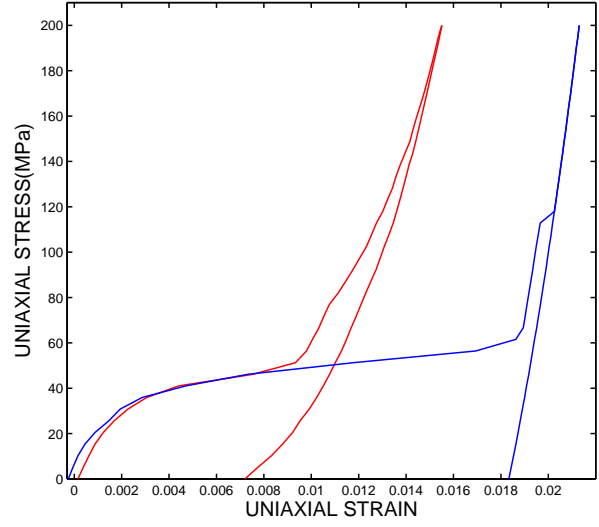}
\caption{
Variation of applied load $\sigma$ with average uniaxial strain
 $\langle\epsilon_{xx}\rangle$ for 
$A_2=-2$ GPa for a polycrystal (red curve) and a single crystal (blue curve). 
}
\label{fig1}
\end{figure}
\begin{figure}[h]
\epsfysize=21cm
\epsffile{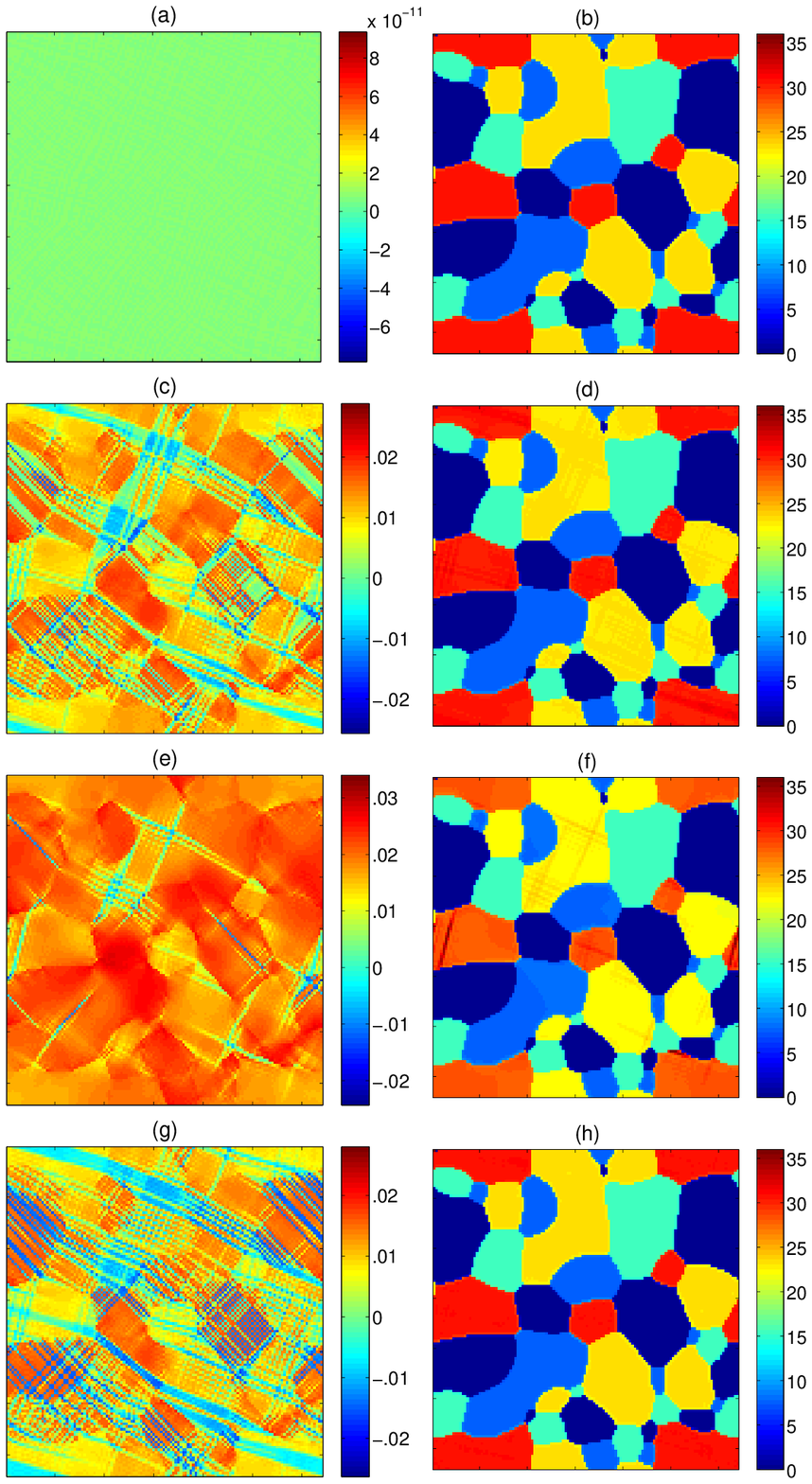}
\caption{
Spatial distribution of $\epsilon_2(\vec{r})$ 
(deviatoric strain in a global frame) for a polycrystal with $A_2=1$ GPa
(snapshots (a),(c),(e) and (g)) and $\theta(\vec{r})$
(snapshots (b),(d),(f) and (h)). The corresponding stress levels are
$\sigma=0$ (before loading, (a) and (b)),
$\sigma=30.76$ MPa ((c) and (d)),
$\sigma=200$ MPa ((e) and (f)) and
$\sigma=0$ (after unloading, (g) and (h)).
}
\label{fig1}
\end{figure}
\newpage
\begin{figure}[h]
\epsfysize=21cm
\epsffile{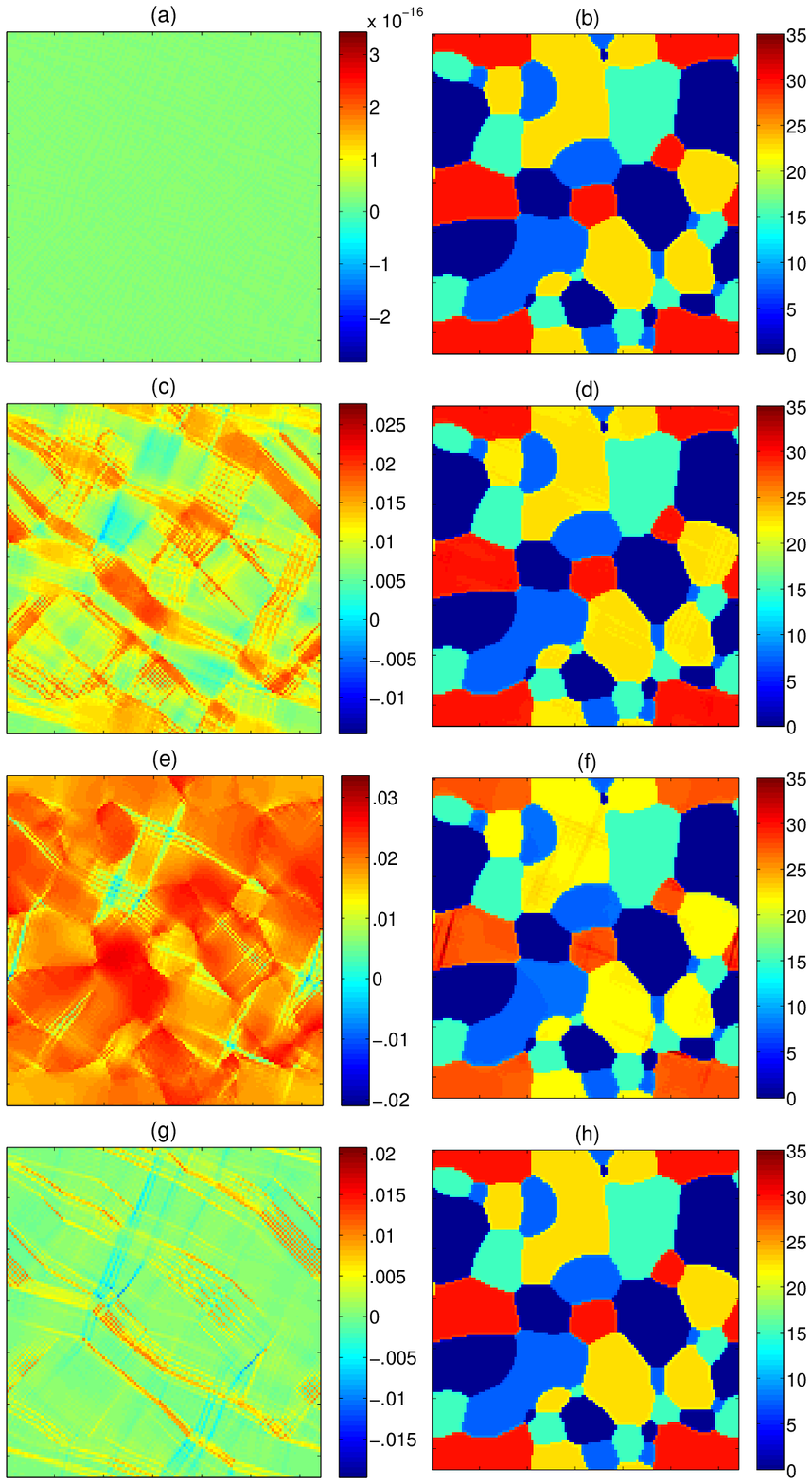}
\caption
{Spatial distribution of $\epsilon_2(\vec{r})$ 
(deviatoric strain in a global frame) for a polycrystal with $A_2=2$ GPa
(snapshots (a),(c),(e) and (g)) and $\theta(\vec{r})$ in degrees
(snapshots (b),(d),(f) and (h)). The corresponding stress levels are
$\sigma=0$ (before loading, (a) and (b)),
$\sigma=30.76$ MPa ((c) and (d)),
$\sigma=200$ MPa ((e) and (f)) and
$\sigma=0$ (after unloading, (g) and (h)).}
\label{fig1}
\end{figure}
\begin{figure}[h]
\epsfysize=21cm
\epsffile{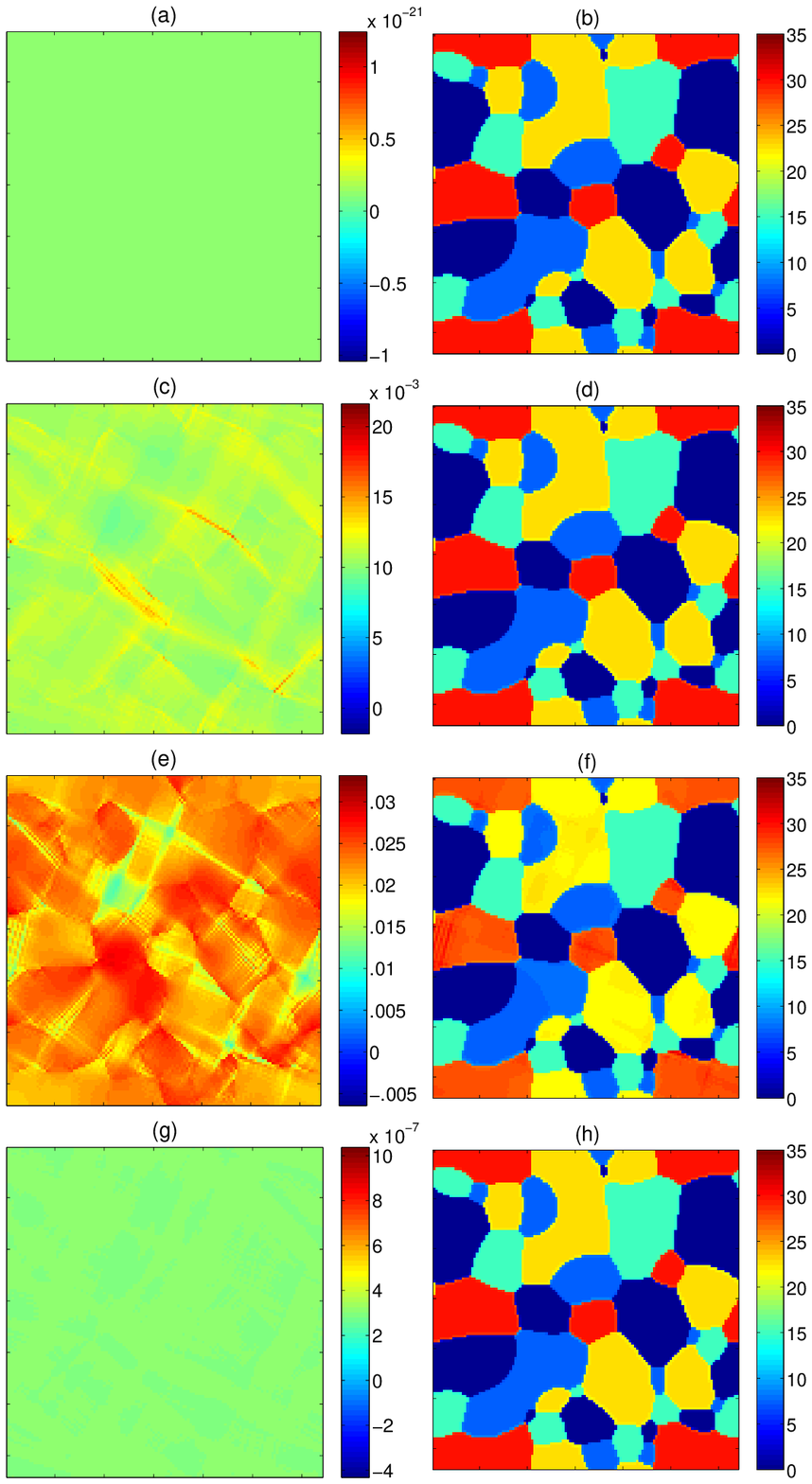}
\caption
{Spatial distribution of $\epsilon_2(\vec{r})$ 
(deviatoric strain in a global frame) for a polycrystal with $A_2=3$ GPa
(snapshots (a),(c),(e) and (g)) and $\theta(\vec{r})$ in degrees
(snapshots (b),(d),(f) and (h)). The corresponding stress levels are
$\sigma=0$ (before loading, (a) and (b)),
$\sigma=30.76$ MPa ((c) and (d)),
$\sigma=200$ MPa ((e) and (f)) and
$\sigma=0$ (after unloading, (g) and (h)).
}
\label{fig1}
\end{figure}
\newpage
\begin{figure*}
\epsfysize=20cm
\epsffile{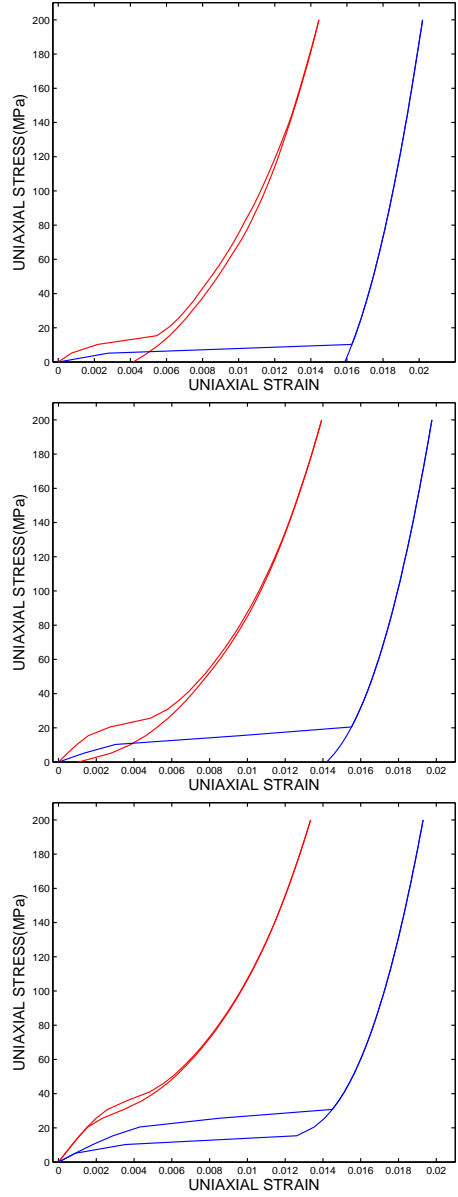}
\caption{ Variation of applied load $\sigma$ vs. average 
uniaxial strain $\langle \epsilon_{xx} \rangle$ as a function of 
(increasing) temperature. From top to bottom,
$A_2=1, 2, 3$ GPa, respectively. Red curves correspond to a polycrystal and blue
curves to a single crystal.}
\end{figure*}
\end{document}